\documentclass[%
 preprint, 
 amsmath,amssymb,
 aps, physrev,
]{revtex4-2}

\usepackage{siunitx}
\usepackage{graphicx,amsmath}
\usepackage{dcolumn}
\usepackage{bm}
\usepackage{braket} 

\begin{document}

\preprint{APS/123-QED}

\title{\textbf{Pervasive Vulnerability Analysis and Defense for QKD-based Quantum Private Query}} 

\author{Xiaoyu Peng}
\author{Bin Liu}
\email{binliu@cqu.edu.cn}
\author{JiaHao Huang}
\author{Nankun Mu}
\affiliation{College of Computer Science, Chongqing University, Chongqing 400044, China.}

\author{Wei Huang}
\author{Bingjie Xu}
\affiliation{Science and Technology on Communication Security Laboratory, Institute of Southwestern Communication, Chengdu 610041, China} 
\author{Fei Gao}
\affiliation{State Key Laboratory of Networking and Switching Technology, Beijing University of Posts and Telecommunications, Beijing 100876, China}


\date{\today}

\begin{abstract}
Quantum Private Query (QPQ) based on Quantum Key Distribution (QKD) is among the most practically viable quantum communication protocols, with application value second only to QKD itself. 
However, prevalent security vulnerabilities in the post-processing stages of most existing QKD-based QPQ protocols have been severely overlooked.
This results in the database being far less secure against individual attacks than indicated by previous protocol analyses.
This study focuses on hidden information extraction under undetermined signal bits, revealing that most such QPQ protocols face severe security threats even without complex quantum resources. Specifically, direct observation attack causes incremental information leakage, while the minimum error discrimination attack efficiently steals additional database inforamtion. 
To address these critical flaws, the proposed multi-encryption defense scheme is compatible with existing QPQ protocols. 
The study demonstrates the necessity of the multi-encryption strategy for the security of databases in QPQ, providing key theoretical and technical support for constructing practical QPQ protocols resistant to real-world attacks.
\begin{description}
\item[Usage]
The findings are intended for three applications: evaluating security vulnerabilities in the pioneering QKD-based QPQ protocol, providing a readily deployable defense scheme leveraging multi-layer encryption keys, and guiding the design of more robust future protocols.
\item[Structure]
The paper is organized as follows: Section II introduces the background and attack models. Sections III and IV analyze the two proposed attacks. Section V introduces multi-encryption defense strategy. Section VI concludes the work.
\end{description}
\end{abstract}

\maketitle


\section{Introduction and Research Motivation}

Quantum information science offers powerful tools for enhancing communication security, with quantum key distribution (QKD) guaranteeing unconditional security grounded in the fundamental principles of quantum mechanics\cite{Gisin2002,Scarani2009}. The seminal BB84 protocol\cite{Bennett1984} laid the theoretical and technical foundation for practical quantum cryptography, and subsequent protocols such as the SARG04 protocol\cite{Scarani2004} have since become the core underpinnings of QKD-based quantum private query (QPQ) systems.

Against this technical backdrop, QPQ has emerged as a highly promising cryptographic primitive that enables mutual privacy protection between communicating parties\cite{Gertner2000,Lo2005}.
QPQ protocols can be broadly categorized into two types: quantum-computation-based QPQ \cite{G08,E11} and QKD-based QPQ. The former is characterized by lower communication complexity, while the latter provides stronger protection for user privacy and is more feasible to implement under current technical conditions.
Derived from the SARG04 protocol, Jakobi et al. proposed the first QKD-based QPQ protocol in 2011\cite{Jakobi2011}, which is denoted as the J-protocol. This protocol enables users to perform database queries while preserving the privacy of both the database contents and the query itself.
It is noteworthy that QKD-based QPQ protocols, including the J-protocol, fall into the category of secure computation protocols; as such, unconditionally secure private queries have been proven impossible\cite{nogolo97}. Consequently, QPQ protocols are designed to achieve a notable level of security rather than perfect security. This work does not claim to have identified any flaws in the unconditional security of such protocols, but instead provides a more precise quantification of the actual security level attainable with QKD-based QPQ protocols. However, existing security analyses have largely overlooked the critical inherent vulnerabilities in the post-processing phase of these protocols.

QKD-based QPQ protocols utilize QKD technology to generate an oblivious key. Generally, such protocols can be divided into the following three stages:
\begin{enumerate}
  \item Raw oblivious key distribution: In this stage, the database and the user generate a raw oblivious key using QKD technology, which satisfies the following two conditions: (1) the database possesses complete information about the raw oblivious key and (2) for each bit of the raw oblivious key, the user knows its value with a certain probability.
  \item Post-processing: In this stage, the database and the user jointly process the raw oblivious key into the final oblivious key, which meets the following conditions: (1) the database has full information of the final oblivious key, (2) the user knows slightly more than one bit of the final oblivious key, typically 2 to 5 bits, and (3) the database does not know which bits the user gains.
  \item Private query: With the final oblivious key in place, the database and the user execute the private query task.
\end{enumerate}

The vast majority of current QKD-based QPQ protocols are derived from the J-protocol and adopt analogous post-processing workflows. 
For instance, Yang et al. proposed a QPQ protocol based on the B92 protocol; Zhao et al. proposed a QPQ protocol based on the measurement-device-independent (MDI) QKD protocol; Yu et al. designed a QPQ protocol with a communication structure opposite to that of the J-protocol based on the SARG04 protocol, aiming to better protect user privacy. 

These protocols leverage quantum properties during the initial oblivious key distribution stage, resulting in a probability $p$ of successful signal decoding by the user, whereas the signal content remains indeterminate with a complementary probability of $1-p$. 
Specifically, in the J-protocol and the protocol proposed by Yu et al., $p = 1/4$; in the QPQ protocols based on B92 and MDI QKD, $p = \sin^2\theta/2$, where $\theta$ is the parameter of the encoded state in the protocol. 

Prior security analyses have rarely conducted systematic investigations into user signal processing behaviors under such indeterminate conditions\cite{Wei2012,Gao2014}. In contrast, our work demonstrates that from the user's perspective, these indeterminate signals follow a biased distribution, which is a characteristic that can give rise to unintended leakage of database information. The high degree of structural similarity in post-processing across different QPQ implementations renders attack strategies targeting this phase universally applicable, thereby posing a severe threat to the overall security of QKD-based QPQ protocols.

Given the J-protocol's foundational role in the development of QPQ technology, this paper selects it as a representative case study to systematically analyze potential database information leakage in the $1-p$ indeterminate scenario, with a specific focus on attack strategies that malicious users can exploit to compromise database privacy. 
We further investigate how vulnerabilities in the J-protocol's post-processing phase can be exploited, and subsequently propose robust defense mechanisms to enhance database security against such malicious attacks—thus advancing QPQ technology toward practical, high-security deployment. 
Owing to the structural similarities in post-processing and key distillation steps across different QKD-based QPQ protocols, the attack methods developed for the J-protocol can be readily extended to other protocols within this category.

\section{Overview of J-protocol}

In a typical J-protocol scheme, the communication parties, the user (Alice) and the database server (Bob), interact according to the following steps:

\begin{enumerate}
  \item Bob prepares and sends the randomly chosen quantum states from the set $\{ \ket{0}, \ket{1}$,
$\ket{+},\ket{-} \}$ to Alice.
  \item Alice randomly selects either the $Z$ basis $ \{ \ket{0},\ket{1} \}$ or the $X$ basis $ \{ \ket{+},\ket{-} \}$  to measure each quantum state that enters. If her measurement is successful (yielding a definite result), she sends an acknowledgment signal to Bob.
  \item Upon receiving the acknowledgment, Bob declares two states for the corresponding signal: the actual state he sent and another state chosen randomly from the remaining state in the complementary basis.
  \item In this process, Alice can deterministically identify the prepared state with a probability of 1/4, while Bob cannot determine which state Alice has successfully identified.
  \item After generating a raw key string, a key distillation step is performed. In the classical post-processing phase, the raw key is compressed by dividing it into sub-strings and adding them bitwise. This ensures that, probabilistically, Alice can only deterministically know a limited number of bit positions in the final key. In the ideal case, Alice would know exactly one bit, but in practice, to ensure protocol functionality, she may know slightly more than one bit. Alice then informs Bob to cyclically shift his key string so that one of her known bits aligns with the index of the database item she wishes to query. Bob encrypts his database using the shifted key and transmits the ciphertext to Alice. Consequently, Alice can decrypt only the desired item, while excessive leakage of database information is prevented under normal operation.
\end{enumerate}

Having outlined the J-protocol, we proceed to design two distinct attack strategies aimed at compromising its security:

\begin{itemize}
  \item Direct Observation Attack: In this attack, the user is entirely passive and does not deviate from the prescribed protocol. The adversary simply collects the quantum states, performs immediate measurements, and subsequently applies classical post-processing techniques to extract extra information from the inconclusive results. Although its efficiency is limited, this attack poses an immediate threat as it is feasible with current technology.
  \item Minimum-Error Discrimination Attack: In contrast to the passive approach, this attack involves an active strategy where the user employs optimal quantum measurements to distinguish between non-orthogonal states with the minimum possible error. This sophisticated method achieves higher efficiency but requires advanced capabilities such as quantum memory, representing a more demanding technological challenge.
\end{itemize}

The attack analysis in this paper primarily targets the malicious user scenario. The direct observation attack can be executed by honest-but-curious users (requiring no additional quantum resources), while the MED attack requires malicious user capabilities (such as quantum memory and non-standard measurement equipment).

The direct observation attack can be directly applied to most QKD-based QPQ protocols. In contrast, the MED attack requires additional steps tailored to the communication structure of different protocols before it can be reduced to the attack framework developed for the J-protocol in this paper. For instance, in the protocol proposed by Yu et al. and the MDI QKD-based protocol by Zhao et al., the malicious user can prepare a Bell state, involve one particle in the protocol process, retain the other particle, and subsequently perform an MED attack on the retained particle based on the measurement information obtained from the database.

\section{Direct Observation Attack}

This section analyzes the amount of database information accessible to a user using the Honest-but-Curious (HbC) attack model. In this model, the user strictly adheres to all steps specified by the protocol, without any active deviation, message tampering, or refusal to cooperate, thus appearing fully compliant. However, the HbC attacker records all accessible intermediate data and communication information throughout the process, attempting to infer additional database content by analyzing this information afterwards.

Taking the J-protocol as an example, the encoding rule is defined as follows: the basis $Z$
($\{\ket{0},\ket{1}\}$)
represents bit 0, while the basis $X$
($\{\ket{+},\ket{-}\}$)
represents bit 1. 
Analysis shows that for each quantum state sent by the server, the user can only deterministically identify the corresponding state with a probability of $1/4$. And there is a $3/4$ probability that the user cannot determine the exact value of the key bit.

All previous protocols assume that when the user cannot determine the exact value of the key bit, they do not gain any information. However, our findings indicate that even in such cases, significant correlation information can still be obtained through probabilistic calculations. 
For cases where the user Alice cannot obtain a definite bit value, we take the scenario where Alice's measurement result is $|+\rangle$ and the database Bob publishes $\{|0\rangle, |+\rangle\}$ as an example. 
Before Alice performs the measurement, the probabilities that Bob encodes 0 (using the $Z$ basis) and 1 (using the $X$ basis) are $P(Z) = P(X) = 1/2$, respectively. Considering the probabilities of the measurement result $|+\rangle$ under different encoding conditions of Bob, we have $P(|+\rangle|Z) = 1/2$ and $P(|+\rangle|X) = 1$. 
Then, the total probability of obtaining the measurement result $|+\rangle$ is calculated as 
\begin{equation}
P(|+\rangle) = P(|+\rangle|Z)P(Z) + P(|+\rangle|X)P(X) = \frac{3}{4}.
\end{equation}
At this point, according to Bayes' theorem, the posterior probabilities are
\begin{equation} 
P(Z||+\rangle) = P(|+\rangle|Z)P(Z)/P(|+\rangle) = 1/3
\end{equation}
and
\begin{equation}
P(X||+\rangle) = P(|+\rangle|X)P(X)/P(|+\rangle) = 2/3. 
\end{equation}
In other words, the probabilities that Bob encodes the bit 0 and bit 1 are 1/3 and 2/3, respectively.
In other words, when the protocol itself does not allow for determining the exact value of a specific raw key bit, the measurement result has a probability $2/3$ of being consistent with the sent state. This subtle probabilistic bias was not considered a serious information leak in the original security analysis. Nevertheless, our calculations reveal that even after the dilution step designed to reduce the additional information, multiple queries can still lead to the acquisition of a substantial amount of additional information, thereby causing significant leakage of database information.

In light of the above facts, we present \textbf{the attack strategy of the Honest-but-Curious user}: when a certain key bit is uncertain, set its value equal to the bit represented by the measurement basis chosen by the user.
Next, we proceed to analyze the effectiveness of this attack.
For the clarity of description and reader understanding, we hereby introduce two probability-related definitions: 
\begin{equation}
P_{j} =\binom{k}{j}\left(\frac{1}{4}\right)^{k-j}\left(\frac{3}{4}\right)^{j},
\end{equation}
which denotes the probability of the event that a user has exactly $j$ uncertain bits out of $k$ raw key bits, and 
\begin{equation}
Q_j = \sum_{\substack{i = 0}}^{\left \lfloor \frac{j}{2} \right \rfloor }\binom{j}{2i}\left(\frac{2}{3}\right)^{j-2i}\left(\frac{1}{3}\right)^{2i}
\end{equation}
which denotes the probability that the user acquires correct database information under the condition that the event corresponding to $P_j$ happens.
Thus, for each final key bit in a single-round query, the average amount of information obtainable by the HbC user is
\begin{equation}
I_s=\sum_{j=0}^kP_j(1-H(Q_j)),
\end{equation}
where $H(p) = -p \log_2 p - (1-p) \log_2 (1-p)$.
Table \ref{tab:HbCs} presents a comparison of the information obtainable by the HbC user and the honest user under different values of $k$ in the J-protocol. 
It can be observed that as the value of $k$ increases, the advantage gained by the HbC user also increases significantly.
\begin{table}[b]
\caption{\label{tab:HbCs}
A comparison of the average information obtainable per key bit between the HbC user and the honest user}
\begin{ruledtabular}
\begin{tabular}{c c c c c c c}  
\multicolumn{1}{c}{\textrm{$k$}} &
\multicolumn{1}{c}{3} &
\multicolumn{1}{c}{4} &
\multicolumn{1}{c}{5} &
\multicolumn{1}{c}{6} &
\multicolumn{1}{c}{7} &
\multicolumn{1}{c}{8}  \\
\colrule
HbC User         & 0.0313 & 0.0101  & 3.27$\times10^{-3}$ & 1.07$\times10^{-3}$ & 3.49$\times10^{-4}$ & 1.15$\times10^{-4}$\\
Honest User  & 0.0156 & 0.0039 & 9.77$\times10^{-4}$ & 2.44$\times10^{-4}$ & 6.10$\times10^{-5}$ & 1.53$\times10^{-5}$\\
Ratio            & 2.0   & 2.58   & 3.34               & 4.36               & 5.72    & 7.53 \\
\end{tabular}
\end{ruledtabular}
\end{table}

Next, we analyze the scenario of multi-round queries. 
The probability distribution of database items can be obtained from the per-round key probability distribution. 
And the scenario of multi-round queries can also be calculated using conditional probability. 
However, as the number of query rounds increases, the number of possible probability distributions grows exponentially, making it extremely difficult to directly compute the exact results. 
Therefore, we resort to simulation for analysis.

\begin{figure}[htbp]
\centering
\includegraphics[width=0.8\textwidth]{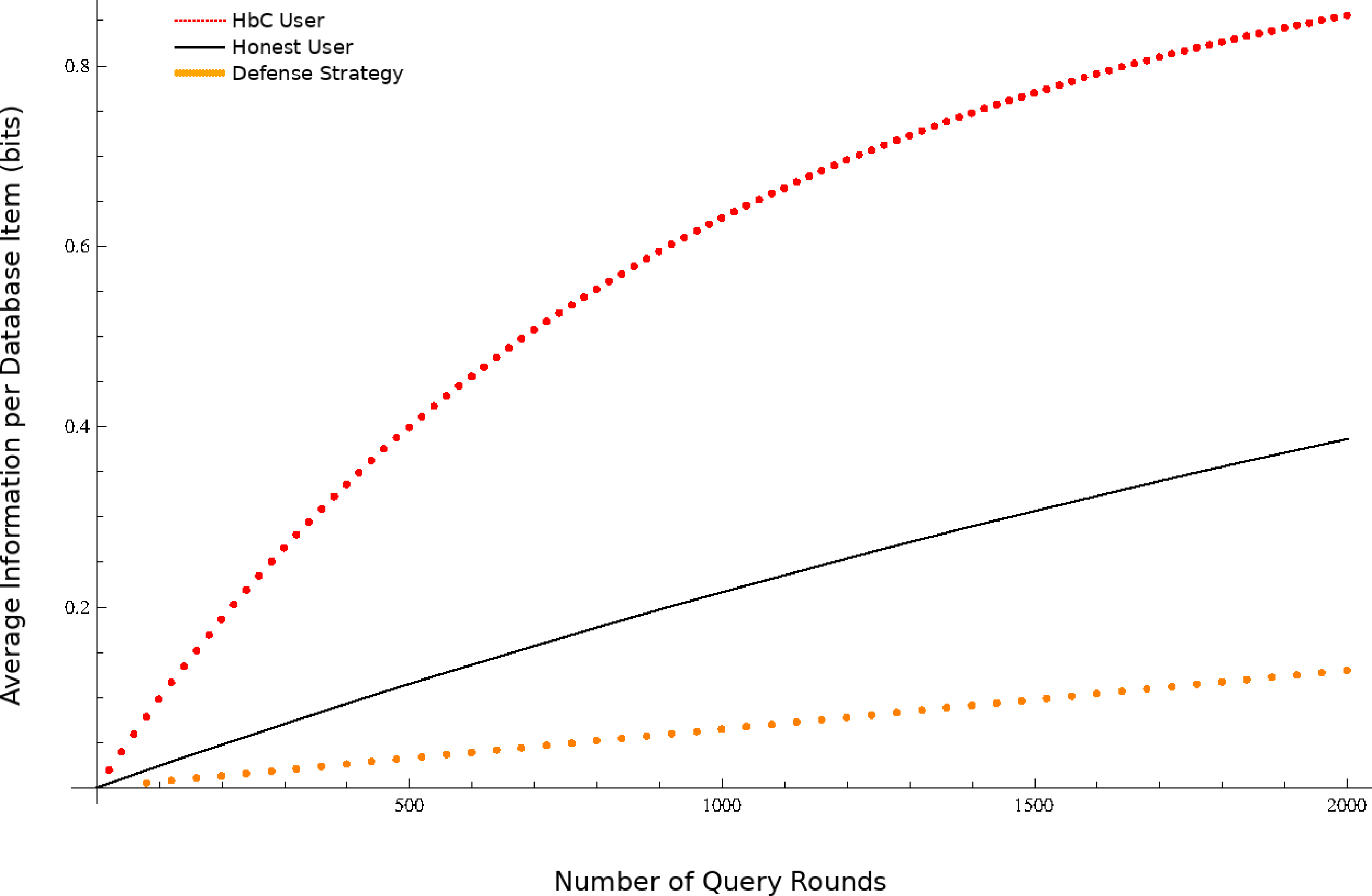} 
\caption{\label{HbCm} Simulation of multi-round attacks by the HbC user when $k=6$. The red (dense dotted) curve depicts the average amount of information obtained by the HbC user for each database item over 2000 rounds of queries, the black curve represents the average amount of information accessible to the legitimate user (which is actually the proportion of database items obtained relative to the entire database), and the orange (sparse dotted) curve represents the effect of the  defense strategy described in Section \ref{S} against HbC attack. Note that the database size needs to be specified in defense simulation, and here we set the database to contain 32000 items.}
\end{figure}

Let $\{P^{(0)}_{i-1},P^{(1)}_{i-1}\}$ denote the probability distribution of a specific database item after the $(i-1)$-th query, and $\{p^{(0)}_{i},p^{(1)}_i\}$ denote the database item distribution from the $i$-th query, where $P^{(0)}_{i-1}+P^{(1)}_{i-1}=1$ and $p^{(0)}_{i}+p^{(1)}_i=1$.
Then the database item distribution after the $i$-th query becomes $\{P^{(0)}_{i},P^{(1)}_{i}\}$,where
\begin{equation}
P^{(0)}_{i}=\frac{P^{(0)}_{i-1}\times p^{(0)}_{i}}{P^{(0)}_{i-1}\times p^{(0)}_{i} +P^{(1)}_{i-1}\times p^{(1)}_{i}},\ \ P^{(1)}_{i}=1-P^{(0)}_{i}.
\end{equation}
For the probability distribution generated in each round of simulation, if the true value of the database item is 0, then $\{p^{(0)}_{i}, p^{(1)}_i\}$ takes the value $\{Q_j, 1-Q_j\}$ with probability $P_j$; if the true value of the database item is 1, then $\{p^{(0)}_{i},p^{(1)}_i\}$ takes the value $\{1-Q_j, Q_j\}$ with probability $P_j$.
Taking $k=6$ as an example (corresponding to a database size ranging from 10,000 to 40,000 in the J-protocol), Fig. \ref{HbCm} shows the average amount of information derived from 100,000 simulated attacks. After 2,000 rounds of queries, the average information per database item obtained by the HbC user reaches 0.856, while that of the legitimate user is only 0.386, corresponding to a ratio of 2.22. It can be observed that compared with single-round queries, the advantage of the HbC user is diluted in multi-round query scenarios.

In fact, if the behavior of malicious users is not restricted, such users can declare a more favorable shift based on the data obtained from previous queries and the key structure of the current round, thereby acquiring database information with higher efficiency.
Meanwhile, multiple malicious users can also collude to claim the most favorable combination of shifts, which is more effective than a single user conducting multiple queries. However, simulating these attack methods is extremely complex even in a simulated environment.
Since we propose a more efficient attack strategy in the next section, we do not conduct further analysis of the aforementioned attack methods here.

\section{Minimum-Error Discrimination Attack}

As described in the previous section, malicious users can adopt more active attack strategies to steal database information with higher efficiency. Here, we consider the scenario where malicious users can temporarily store the received quantum signals. In this way, after the database publishes the state set, malicious users can implement more efficient attack strategies.

Considering each bit in the raw oblivios key, after the database publishes the set of two quantum states, the extraction of database encoding information by the user is transformed into a problem of distinguishing between the two quantum states. 
Previous studies have primarily focused on attacks utilizing unambiguous discrimination (UD). Under such attacks, for each raw key bit, a malicious user can obtain each key bit with a probability of 
\begin{equation}
p_{UD}=1-F(\rho_0,\rho_1)=1-|\langle0|+\rangle|=1-\frac{\sqrt{2}}{2},
\end{equation}
instead of the normal probability of 1/4, where $F(\rho_0,\rho_1)$ is the fidelity between the two quantum states the database declares in Step 3.
Different from the scenario analyzed in the previous section, after adopting the unambiguous discrimination attack, no additional information leakage will occur for the other bits that are not explicitly obtained.

However, our analysis shows that the threat posed by minimum-error discrimination (MED) is far more severe.
It is straightforward to derive that the minimum error probability for a malicious user to distinguish the stored quantum states is
\begin{eqnarray}
p_{\text {e}}  =  \frac{1}{2}\left(1-D(\rho_0,\rho_1)\right) = \frac{1}{2}\left(1-\sqrt{1-|\langle0|+\rangle|^2}\right)=\frac{1}{2}\left(1-\frac{\sqrt{2}}{2}\right), 
\end{eqnarray}
where $D(\rho_0,\rho_1)$ is the trace distance between the two quantum states the database declares in Step 3.
That is, the malicious user has a probability of about 0.8535 to obtain the correct raw key bit.
Then for each final key bit, the probability that the malicious user obtains the correct result is
\begin{eqnarray}
P_{\text{c}}= \sum_{\substack{i = 0}}^{\left \lfloor \frac{k}{2} \right \rfloor }\binom{k}{2i}\left(1-p_{\text {e}}\right)^{k-2i}\left(p_{\text {e}}\right)^{2i}.
\end{eqnarray}
Table \ref{tab:mea} demonstrates the advantages of the MED individual attack over the UD individual attack in a single-round scenario under different values of $k$.
It can be observed that as $k$ increases, the advantages of the MED attack also increase significantly.
\begin{table}[b]
\caption{\label{tab:mea}
A comparison of the average information obtainable per key bit between the users adopting the MED attack and those adopting the UD attack}
\begin{ruledtabular}
\begin{tabular}{c c c c c c c}  
\multicolumn{1}{c}{\textrm{$k$}} &
\multicolumn{1}{c}{3} &
\multicolumn{1}{c}{4} &
\multicolumn{1}{c}{5} &
\multicolumn{1}{c}{6} &
\multicolumn{1}{c}{7} &
\multicolumn{1}{c}{8}  \\
\colrule
MED Attack    & 0.0921 & 0.0456  & 0.0227 & 0.0113 & \num{5.64E-3} & \num{2.82E-3}\\
UD Attack  & 0.0251 & \num{7.36E-3} & 2.16$\times10^{-3}$ & 6.31$\times10^{-4}$ & 1.85$\times10^{-4}$ & 5.41$\times10^{-5}$\\
Ratio            & 3.67   & 6.19   & 10.5              & 17.9              & 30.5    & 52.1 \\
\end{tabular}
\end{ruledtabular}
\end{table}
Next, we consider the scenario of multi-round queries, assuming a total of $m$ rounds.
For each item of database, an overestimated target bit value will be obtained in each round of queries, and the value may vary across different rounds. The probability that the target bit value is identical in exactly $j$ out of $m$ ($j> m/2$) rounds is given by
\begin{equation}
P^\prime_j=\binom{m}{j}\left({P_c}^{j}(1-P_c)^{m-j}+{P_c}^{m-j}(1-P_c)^{j}\right).
\end{equation}
In this scenario, the overall probability bias of the database item obtainable by the malicious user is
\begin{equation}
Q^\prime_j=\frac{{P_c}^j(1-P_c)^{m-j}}{{P_c}^j(1-P_c)^{m-j}+{P_c}^{m-j}(1-P_c)^{j}}.
\end{equation}
Note that since $j>m/2$ and $P_c>1/2$, $Q^\prime_j>1/2$ and for the case where $j=m/2$, the amount of information obtained by the user is zero.
Thus, the average amount of information per database item obtained via the MED attack is
\begin{equation}
I^\prime_s=\sum_{j=\lceil\frac{m}{2}\rceil}^{m}P^\prime_j(1-H(Q^\prime_j)).
\end{equation}

\begin{figure}[htbp]
\centering
\includegraphics[width=0.8\textwidth]{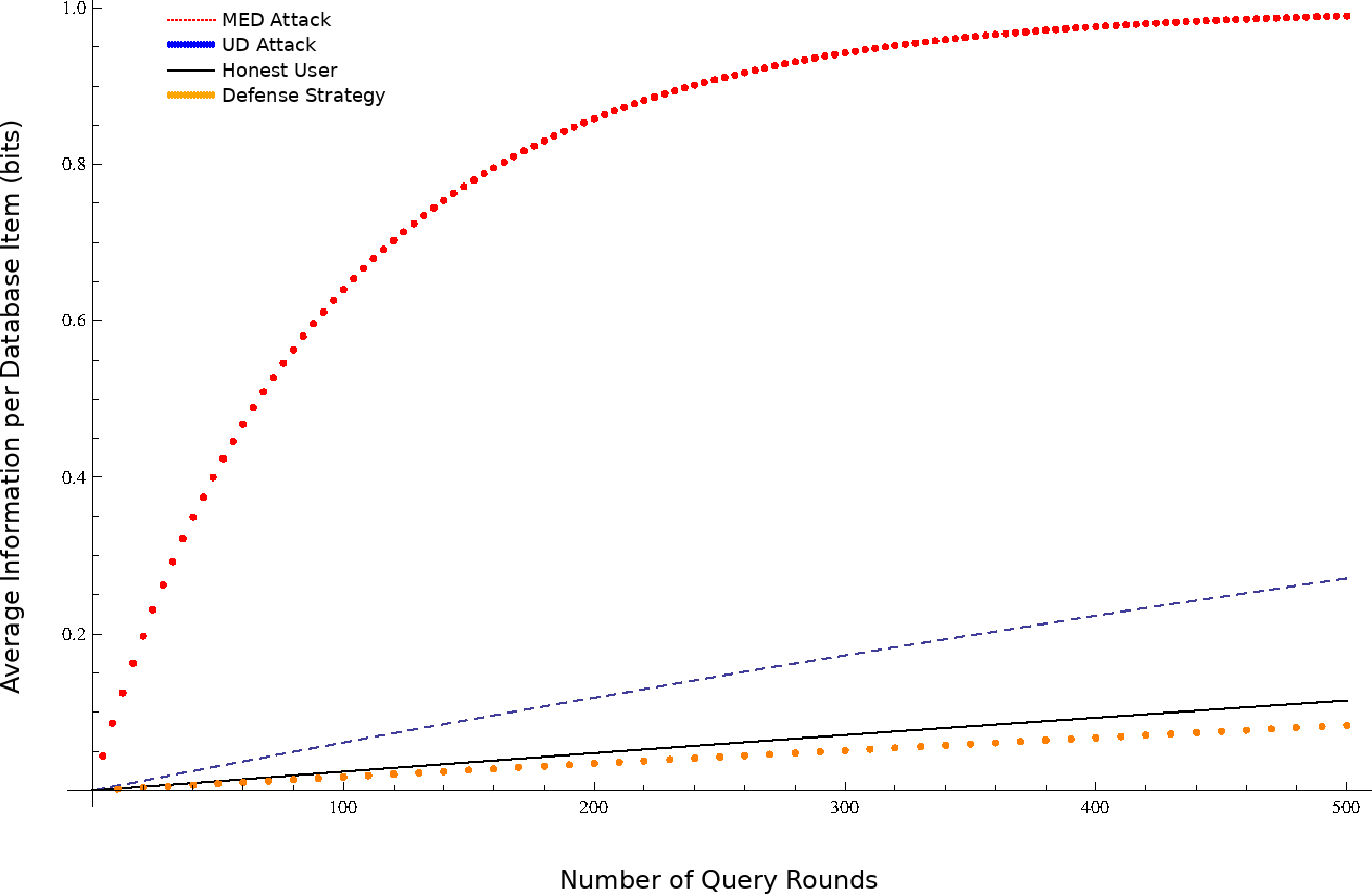} 
\caption{\label{fmedm} A comparison of the effectiveness between the MED individual attack  proposed in this paper and the conventional UD individual attack. Herein, the red (dense dotted) curve represents the efficiency of the MED attack, the blue (dashed) curve represents that of the UD attack, the black curve represents that of the honest user, and the  orange (sparse dotted) curve represents the effect of the defense strategy described in Section \ref{S}.}
\end{figure}

Likewise, we take the case of $k=6$ (corresponding to a database size ranging from 10,000 to 40,000) as an example. Fig. \ref{fmedm} presents a comparison of the database information obtainable by malicious participants using the MED attack and the UD attack, from which it can be observed that the effectiveness of the MED attack is far superior to that of the UD attack.
\begin{table}[b]
\caption{\label{tab:pac}Average Information $I^\prime_s$ and Average Confidence Probability $P_{ACP}$ with number of MED attack rounds $m$ for k=6 (corresponding to database sizes from 10,000 to 40,000)}
\begin{ruledtabular}
\begin{tabular}{c c c c c c c}  
\multicolumn{1}{c}{\textrm{Round $m$}} &
\multicolumn{1}{c}{1} &
\multicolumn{1}{c}{100} &
\multicolumn{1}{c}{200} &
\multicolumn{1}{c}{300} &
\multicolumn{1}{c}{400} &
\multicolumn{1}{c}{500}  \\
\colrule
$I^\prime_s$  & 0.0113 & 0.640  & 0.858 & 0.942 & 0.976 & 0.990\\
$P_{ACP}$ & 0.5625 & 0.913 & 0.968 &0.987 & 0.995 & 0.998  \\
\end{tabular}
\end{ruledtabular}
\end{table}

To more intuitively demonstrate the actual effectiveness of the MED attack, we list in Table \ref{tab:pac} the confidence probability of the query result for each database entry
\begin{equation}
P_{ACP}=\sum_{j=\lceil\frac{m}{2}\rceil}^{m}P^\prime_jQ^\prime_j,
\end{equation}
and the corresponding average amount of information obtained by the malicious user for each database entry. It can be observed that the average confidence probability exceeds 90\% when the number of queries reaches 100, and surpasses 99\% when the number of queries hits 400.
Thus, in multi-round queries, although a malicious user cannot accurately obtain arbitrary database entries, the emergence of a high confidence probability for database entries as the number of queries increases also constitutes a severe leakage of database information.

As can be observed from Fig. \ref{fmedm}, even when targeting a single-bit individual attack with the MED method, the threat imposed on the QPQ protocol remains extremely significant. 
For example, for a database size ranging from 10,000 to 40,000 items, a malicious user can acquire 99\% of the database information by launching an MED individual attack with merely 500 queries.
In contrast, nearly 8,000 queries are required to launch a UD individual attack to achieve the same effect.
This constitutes a major security loophole that has been overlooked in all previous related studies.

\section{Multi-Encryption Defense Strategy}\label{S}

In recent studies on QPQ protocols with non-ideal light sources, vulnerabilities in database security are typically mitigated by performing multiple encryptions of the database within a single query \cite{decoyQPQ2022,tifs2025}. This strategy is also applicable to defending against the two attack types proposed in this work.
The core idea is that Bob and Alice jointly generate $L$ sets of independent final oblivious keys in a single query; generally, setting $L$ to 2 or 3 is sufficient to reduce the malicious user's advantage to a negligible range. Alice applies a shift to each key set, aligning her known key bits with the target database entry. Bob then encrypts the database data iteratively using these shifted keys, after which the encrypted data is sent to Alice. She decrypts the data layer by layer to retrieve the desired query result.

For the original J-protocol with single-layer key encryption, the protocol execution fails if the user has no known bits in the generated final key, with a failure probability of
\begin{equation}
p_f=(1-\frac{1}{4^k})^N,
\end{equation}
where $N$ is the size of the database.
After adopting the above defense strategy, the failure probability of honest participants in executing the protocol increases from $p_f$ to $1-(1-p_f)^L$, which is less than $Lp_f$.
t is noteworthy that directly increasing the value of k for the single-round key can also reduce the malicious user’s advantage, but this will lead to a substantial increase in the protocol’s failure probability. For instance, instead of encrypting using $L$ final key sets, if the value of $k$ for the single final key is set to $kL$, the failure probability of the protocol will increase to
\begin{equation}
p_f=(1-\frac{1}{4^{kL}})^N.
\end{equation}
Taking a database of size 30,000 as an example, the parameter k=6 is selected in accordance with the J-protocol. The failure probability of the single-layer encryption scheme is 0.066\%, and that of the two-layer encryption scheme is 0.13\%. In contrast, the failure probability of the single-layer encryption scheme with a doubled k reaches 99.8\%; even when k is merely increased to 7, the failure probability still rises to 16\%.

For entries other than the target entry, the probability for the user of successfully decrypting an additional entry is $\eta^L$, where $\eta$ is the single-layer success probability. Compared with the linear increase in the failure probability, the user’s advantage in other entries decreases exponentially. For the two types of attacks mentioned in this paper, this is equivalent to a multiple increase in the protocol parameter $k$ while the database size remains unchanged, thereby leading to a significant reduction in the corresponding advantage of malicious users. When $L=2$, as shown by the orange curve in Figs. \ref{HbCm} and \ref{fmedm}, adopting the multi-encryption defense scheme with two sets of the final oblivious key in each round of query achieves a remarkable defense effect against both the HbC attack and the MED attack. The database information accessible to malicious users is comparable to that obtained by honest users in the protocol using single final key encryption.

It should be noted that since QPQ cannot achieve perfect security, the multi-encryption defense strategy inevitably introduces certain trade-offs. Specifically, the failure probability of honest participants in executing the protocol increases from $p_f$ to $1-(1-p_f)^L$. Additionally, the communication overheads increase by a factor of $L$. Finally, since the user shifts each key set, the opportunity for a malicious database to steal user privacy also increases accordingly. These trade-offs represent the cost of enhanced database security, and the choice of $L$ and $k$ should be determined based on the specific security requirements of the application scenario.

\section{Conclusion}

In the research field of quantum private query (QPQ), except for the joint measurement attack\cite{Liu2025}, scholars generally hold the view that the unambiguous discrimination (UD) attack poses the greatest threat to the database. However, our analysis reveals that from an information-theoretic perspective, the minimum-error discrimination (MED) attack is significantly more threatening than the UD attack, both in single-query and multi-round scenarios, thereby posing a severe risk to database security. Specifically, when $k$ exceeds 5, the efficiency of the MED attack in a single query can be dozens of times that of the UD attack. Furthermore, our confidence probability analysis demonstrates that although the MED attack cannot precisely locate the correctness of each bit, it can extract database information quantity beyond the protocol's design expectations through statistical advantage, validating the actual threat of this attack. Therefore, except for a very few specific protocols\cite{JMQPQ,withoutfQPQ}, the multi-encryption defense scheme should serve as an indispensable step for most QKD-based QPQ protocols. The conclusions drawn in this paper hold significant reference value for subsequent theoretical research on QPQ.

\begin{acknowledgments}
This work was supported by the National Natural Science Foundation of China (Grant No. 62472052).
\end{acknowledgments}

\nocite{*}

\bibliography{refqpq}

@CONTROL{REVTEX42Control}

@CONTROL{apsrev42Control,author="08",editor="1",pages="0",title="0",year="1"}

@article{nogolo97,
	title = {Insecurity of quantum secure computations},
	volume = {56},
	issn = {2469-9926},
	doi = {10.1103/PhysRevA.56.1154},
	number = {2},
	journal = {PHYSICAL REVIEW A},
	author = {Lo, HK},
	month = aug,
	year = {1997},
	pages = {1154--1162},
}

@article{tifs2025,
	title = {Measurement-{Device}-{Independent} {Quantum} {Private} {Query} {With} {Weak} {Coherent} {Source}},
	volume = {20},
	issn = {1556-6013},
	doi = {10.1109/TIFS.2025.3607259},
	journal = {IEEE TRANSACTIONS ON INFORMATION FORENSICS AND SECURITY},
	author = {Liu, B and Yang, TT and Huang, W and Wei, CY and Mu, NK and Xu, BJ and Gao, F},
	year = {2025},
	pages = {9454--9462},
}

@article{E11,
	title = {Secure quantum private information retrieval using phase-encoded queries},
	volume = {84},
	issn = {2469-9926},
	doi = {10.1103/PhysRevA.84.022313},
	number = {2},
	urldate = {2011-08-08},
	journal = {PHYSICAL REVIEW A},
	author = {Olejnik, Lukasz},
	month = aug,
	year = {2011},
}

@article{G08,
	title = {Quantum private queries},
	volume = {100},
	issn = {0031-9007},
	doi = {10.1103/PhysRevLett.100.230502},
	number = {23},
	urldate = {2008-06-13},
	journal = {PHYSICAL REVIEW LETTERS},
	author = {Giovannetti, Vittorio and Lloyd, Seth and Maccone, Lorenzo},
	month = jun,
	year = {2008},
	pages = {230502},
}

@article{decoyQPQ2022,
	title = {Decoy-state method for quantum-key-distribution-based quantum private query},
	volume = {65},
	issn = {1674-7348},
	doi = {10.1007/s11433-021-1843-7},
	number = {4},
	urldate = {2022-03-11},
	journal = {SCIENCE CHINA-PHYSICS MECHANICS \& ASTRONOMY},
	author = {Liu, Bin and Xia, Shuang and Xiao, Di and Huang, Wei and Xu, Bingjie and Li, Yang},
	month = apr,
	year = {2022},
	pages = {240312},
}

@article{Gisin2002,
  author  = {Gisin, N. and Ribordy, G. and Tittel, W. and Zbinden, H.},
  title   = {Quantum cryptography},
  journal = {Reviews of Modern Physics},
  volume  = {74},
  number  = {1},
  pages   = {145--195},
  year    = {2002},
  doi     = {10.1103/RevModPhys.74.145}
}

@article{Scarani2009,
  author  = {Scarani, V. and Bechmann-Pasquinucci, H. and Cerf, N. J. and Dušek, M. and Lütkenhaus, N. and Peev, M.},
  title   = {The security of practical quantum key distribution},
  journal = {Reviews of Modern Physics},
  volume  = {81},
  number  = {3},
  pages   = {1301--1350},
  year    = {2009},
  doi     = {10.1103/RevModPhys.81.1301}
}

@article{Bennett1984,
  author  = {Bennett, C. H. and Brassard, G.},
  title   = {Quantum cryptography: Public key distribution and coin tossing},
  journal = {Theoretical Computer Science},
  volume  = {560},
  pages   = {7--11},
  year    = {2014},
  note    = {Original work presented in 1984}
}

@article{Scarani2004,
  author  = {Scarani, V. and Acín, A. and Ribordy, G. and Gisin, N.},
  title   = {Quantum cryptography protocols robust against photon number splitting attacks for weak laser pulse implementations},
  journal = {Physical Review Letters},
  volume  = {92},
  number  = {5},
  pages   = {057901},
  year    = {2004},
  doi     = {10.1103/PhysRevLett.92.057901}
}

@inproceedings{Gertner2000,
  author    = {Gertner, Y. and Ishai, Y. and Kushilevitz, E. and Malkin, T.},
  title     = {Protecting data privacy in private information retrieval schemes},
  booktitle = {Proceedings of the thirtieth annual ACM symposium on Theory of computing},
  pages     = {151--160},
  year      = {1998},
  doi       = {10.1145/276698.276725}
}

@article{Lo2005,
  author  = {Lo, H.-K.},
  title   = {Security of quantum key distribution},
  journal = {IEEE Journal of Selected Areas in Communications},
  volume  = {24},
  number  = {2},
  pages   = {345--350},
  year    = {2006},
  doi     = {10.1109/JSAC.2005.251481}
}

@article{Jakobi2011,
  author  = {Jakobi, M. and Simon, C. and Gisin, N. and Bancal, J.-D. and Branciard, C. and Walenta, N. and Zbinden, H.},
  title   = {Practical private database queries based on a quantum-key-distribution protocol},
  journal = {Physical Review A},
  volume  = {83},
  number  = {2},
  pages   = {022301},
  year    = {2011},
  doi     = {10.1103/PhysRevA.83.022301}
}

@article{Wei2012,
  author  = {Wei, C.-Y. and Wang, T.-Y. and Gao, F.},
  title   = {Practical quantum private query of blocks based on unbalanced quantum key distribution},
  journal = {Physical Review A},
  volume  = {87},
  number  = {2},
  pages   = {022309},
  year    = {2013},
  doi     = {10.1103/PhysRevA.87.022309}
}

@article{Gao2014,
  author  = {Gao, F. and Liu, B. and Wen, Q.-Y. and Chen, H.},
  title   = {Flexible quantum private queries based on quantum key distribution},
  journal = {Optics Express},
  volume  = {20},
  number  = {16},
  pages   = {17411--17420},
  year    = {2012},
  doi     = {10.1364/OE.20.017411}
}

@article{Liu2025,
  author  = {Liu, B. and Huang, J. and Huang, W. and Li, Z. and Mu, N. and Xu, B. and Gong, B.},
  title   = {Analysis and defense against joint-measurement attacks to quantum private query protocols for multi-rounds},
  journal = {Science China Physics, Mechanics \& Astronomy},
  volume  = {68},
  number  = {2},
  pages   = {280311},
  year    = {2025},
  doi     = {10.1007/s11433-024-2803-x}
}

@article{JMQPQ,
  author  = {Wei, C.-Y. and Wang, T.-Y. and Gao, F.},
  title   = {Practical quantum private query with better performance in resisting joint-measurement attack},
  journal = {Physical Review A},
  volume  = {93},
  number  = {4},
  pages   = {042318},
  year    = {2016},
  doi     = {10.1103/PhysRevA.93.042318}
}

@article{withoutfQPQ,
  author  = {Liu, Bin and Gao, Fei and Huang, Wei and Wen, Qiao-Yan},
  title   = {QKD-based quantum private query without a failure probability},
  journal = {Science China-Physics Mechanics \& Astronomy},
  volume  = {58},
  number  = {10},
  pages   = {100301},
  year    = {2015},
  doi     = {10.1007/s11433-015-5713-2}
}

\end{document}